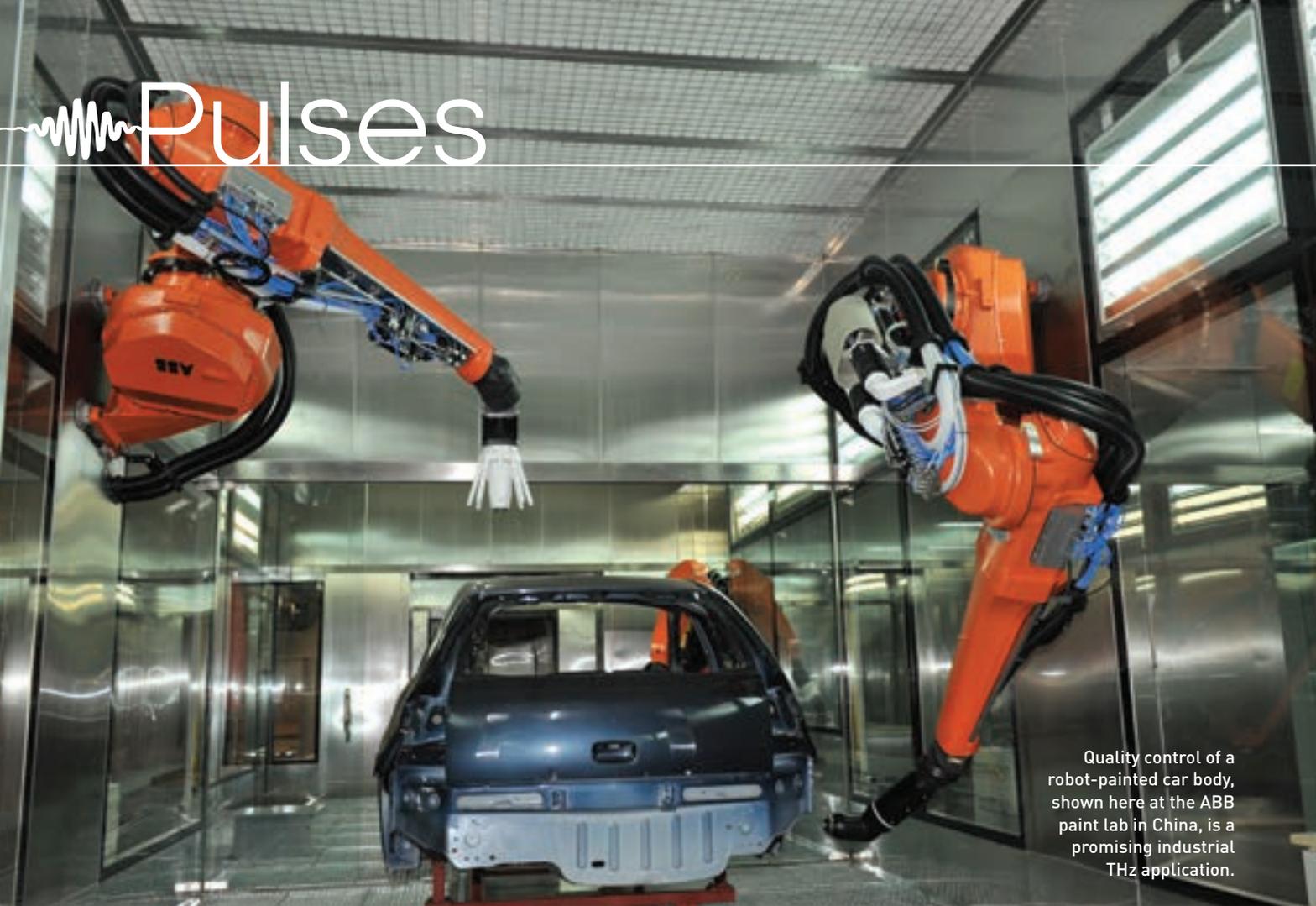

Quality control of a robot-painted car body, shown here at the ABB paint lab in China, is a promising industrial THz application.

Image courtesy of ABB

OPTICS INNOVATIONS

# An Industrial THz Killer Application?

Terahertz technology is mature enough for large-volume sensing applications. However, **Dook van Mechelen** says there are a few hurdles preventing its industrial debut.

Ever since the first laser-induced generation of terahertz (THz) radiation in the 1990s, scientists have enthusiastically embraced THz technology in academic studies. THz spectroscopy has a number of advantages that point to abundant industrial applications as well, in areas such quality control, security and biomedical imaging. Those advantages include detectability at room temperature; a nonionizing nature; applicability to many materials; and, through time-domain techniques, the ability to provide depth sensitivity, nondestructively.

Yet despite those advantages, the search for a THz "killer application"—a novel, innovative use with a business case strong enough to bring the technology into the industrial mainstream—has remained fruitless, and even the hope of finding such an application has begun to falter. Why has a killer app for THz radiation been so elusive? And how can the road to industrial application of this versatile technology be cleared?

## Asking the right questions

It is easy to come up with candidate processes and materials—paper sheets, paint layers, pharmaceutical coatings, plastics—that could offer a suitable testbed for the first industrial-volume THz application. These materials all have comparatively little water content, and are partially transparent, but thick enough to provide a clear optical contrast. Yet the process industry already successfully employs other optical sensors and techniques in these fields. Therefore, the key question to answer is: What can THz technology do cheaper, do more of, or do better than other optical techniques?

One answer to that question is that, for a wide range of materials, THz could determine a variety of quality parameters in a single, non-time-averaged measurement—and enable physical identification of a material well beyond what is known to be THz spectroscopic "fingerprint" identification. That capability, in turn, derives from three key potential elements:

***A dedicated and sufficiently general analysis scheme:*** The THz analysis scheme is the core element for obtaining several parameters at once. The ideal analysis scheme would describe the light-matter interaction as closely as possible, and would be general enough to apply to a wide range of materials.

***Broadband emission:*** Optical sensing is commonly carried out in narrow frequency bands from which the material's optical functions cannot be reconstructed, and thus requires multiple measurements and material-specific signal processing to tease out a single quality parameter. In contrast, THz's large-bandwidth emission, often probing a variety of spectral features, allows reliable deduction of the material's optical functions, thereby increasing the overall accuracy of the sensed parameters.

***Fast time-domain acquisition:*** Time-domain data, acquired in short times to keep up with fast industrial processes, are well-suited for extracting geometrical properties of complicated material structures.

In sum, THz time-domain sensing, empowered by a sufficiently general analysis scheme—while not cheaper than other optical techniques—can do much more. In principle, it can provide access to multiple quality parameters at once, through broadband optical functions, and it can do it better, with a more robust characterization than is possible using empirical signal-processing schemes. And it is, above all, oversimplified analysis schemes that have been the biggest stumbling block toward industrial THz applications.

### The real impediment: Inaccurate analysis

Two different strategies can be followed in moving a technology from the lab to industrial applications: the high-risk approach of creating an entirely new application, and new market, from scratch; and the ostensibly less risky option of

## Above all, oversimplified analysis schemes have been the biggest stumbling block toward industrial THz applications.

### THz for industry: Extracting material properties

New methods for extracting material properties from THz time-domain data are coming to the forefront. Precise results can be obtained through modeling the light-matter interaction. The electric field of the reflected and transmitted rays after interaction with a sample can be calculated using the Fresnel equations and an appropriate dispersion model. Material parameters are then conveniently obtained by fitting to the experimental data.

The graphs below are an example of three fitting analysis schemes with increasingly accurate dipersion models used for a paint layer:

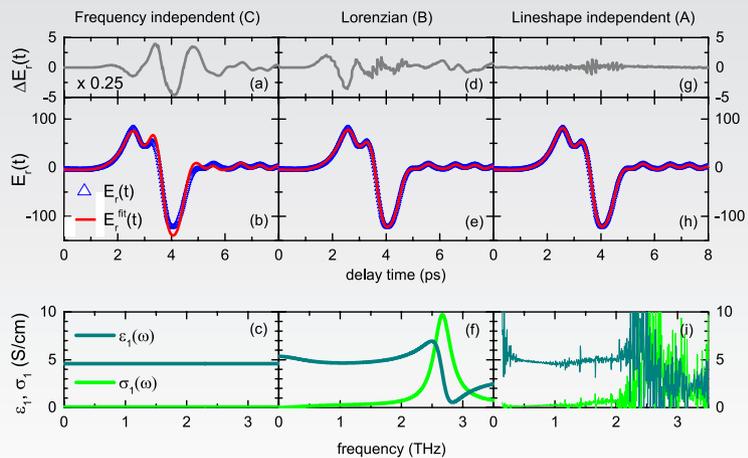

For a detailed illustration of how to obtain material properties from THz time-domain data, go to the online supplement at **www.osa-opn.org/THz_analysis**

> A dedicated analysis scheme has the potential to accurately provide a number of material property parameters essential for industry's strict requirements.

replacing an older, established technology with a newer, superior one in an existing application space. The latter approach could well prove the right one for THz—yet the revamped THz sensor still must outperform the current state-of-the-art, and here, the technological performance of THz has so far fallen short.

For sensing, many believe that the power of existing THz sources constitutes the main limiting factor in the quest for a killer app.

Instead, the real impediment is on the signal-processing side—and, in particular, the oversimplified analysis schemes currently employed to obtain material properties. As already noted, a dedicated analysis scheme has the potential to accurately provide a number of material property parameters essential for industry's strict requirements. Yet current schemes generally fall short, needing material-specific—and often unknown—input, thereby providing only approximate values of geometric thickness and index refraction, rather than identifying material parameters.

True physical identification is only possible when the analysis scheme describes all relevant aspects of the light-matter interaction, ranging from characteristic excitations, polarization and scattering effects, to morphology and geometrical properties in the industrial environment. Recently, a new materials properties methodology—encompassing Fresnel equations for a stratified structure and appropriate dispersion to model the material transfer function, which in turn is fitted to the THz data to provide the material parameters—has been shown to work for paper sheets and paint layers. But such examples in the field of THz applications are rare. (*A more detailed view of taking material properties from THz time-domain data appears in the online supplement to this article.*)

## The route to industry

New technologies like THz sensing cannot exist solely in academia. The governments and corporations that fund basic research want to eventually move these applications into the market. In this way, publicly funded science can, in turn, stimulate economies and improve societal well-being.

Small companies, such as academic spin-offs, are well-suited for early technology development. Their flexible structure and intellectual resources are key assets to advance the technology at its incunabula. To develop an existing THz technology into an industrial application, a good starting point may be incubating the technology in large firms that produce sensing equipment for the end-user industry. Established firms have more resources to connect the new technology to the market.

The typical industrial approach toward new technologies follows a pragmatic path that begins with examining a range of potential applications for compatibility with existing product lines. Subsequently, market studies are conducted to further define the application and related product with the most prosperous near-term future.

This route thus meanders through academic and industrial research, technology and development, market opportunities and business perspectives, with the application maturing in each step before moving on to the next.

For THz sensing applications, the instrumentation base is mature enough to start on this path to industry. However, the generally oversimplified signal processing for THz spectroscopic analysis applications could be an obstacle for implementation. Addressing this issue will not only enhance the overall performance of the technology but will also make it more accessible to a broader range of industries.

Several large technological industries became interested in THz sensing applications during the last decade, often due to a technology push from small companies such as spin-offs or governmental research centers, and are now exploring the route to the industrial application. Ultimately, the THz "killer application" will be the one that balances the extraordinary properties and the traditionally required performance in the best business case. OPN

Dook van Mechelen (dook.vanmechelen@ch.abb.com) is a physicist leading the THz technology integration at ABB Corporate Research, Switzerland.

**To learn more …**
- A.B. Kuzmenko. Rev. Sci. Instr. **76**, 083108 (2005).
- O.S. Ahmed et al. J. Lightwave Technol. **28**, 1685 (2010).
- T. Hochrein. J. Infrared, Millimeter, Terahertz Waves **36**, 235 (2015).
- J.L.M. van Mechelen. Langmuir **30**, 12748 (2014).
- J.L.M. van Mechelen et al. Opt. Lett. **39**, 3853 (2014).

# Extracting Optical Material Properties from THz Time-Domain Data

Dook van Mechelen

## Time-domain peak subtraction

A commonly used scheme to obtain material properties from THz data is time-domain peak subtraction. The method assumes zero absorption $\sigma_1$ and a frequency independent permittivity $\varepsilon_1 = n^2$ and aims at obtaining an optical thickness $nd$. From there, independent knowledge of either $n$ or $d$ is used to obtain the other.

Among the partial reflections in the time-domain response, two are chosen that differ by a double transmission path. Then, the mutual time delay $\Delta t$ is equaled to $2nd/c$, where $c$ is the speed of light in vacuum. **Figure 1** exemplifies this method for a paint layer on a steel substrate. The time delay between the two peaks in the reflected electric field $E_r^{exp}(t)$ provides $d$ by assuming $n$ as shown in Figure 1c.

Generally, peaks in $E_r^{exp}(t)$ overlap due to the width of the initial electric field pulse $E_r^{ref}(t)$ and small values of $nd$ of multiple layers, causing time identification to be troublesome. A solution is to deconvolve $E_r^{ref}(t)$ out of $E_r^{exp}(t)$, thereby eliminating the line shape of $E_r^{ref}(t)$. In practice, this method only works partially and besides showing more clearly the partial reflections it also creates artificial features, as shown by $E_{r,deconv}(t)$ in Figure 1a.

## Fitting the transfer function

More precise results can be obtained through modeling the light-matter interaction. Both the Fresnel equations and ABCD matrix method may be used to calculate the electric field of one or more reflected and transmitted rays after interaction with the sample. The material parameters can then be obtained either by direct inversion of the optical functions or by fitting the transfer function. This approach has evolved from a scheme for a single slab without absorption to a frequency- and time-domain method on complex multilayer structures taking into account the absorption and dispersion of each layer.

**Figure 2** shows the result of three of these analysis schemes based on fitting, applied onto the data of Figure 1. Each successive scheme describes the light-matter interaction more realistically. Eventually, characteristic parameters of the five schemes (labeled A, B, C for the fitting, and D1 and D2 for peak subtraction) are mutually compared in Figure 3.

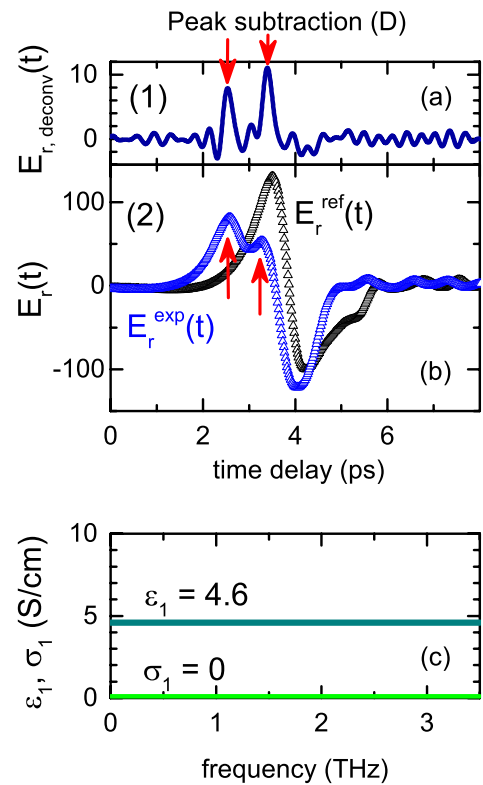

**Figure 1.** Peak subtraction analysis schemes illustrated for THz time-domain data of a paint layer on steel in reflection. (a) The deconvolved sample response $E_{r,deconv}(t)$; (b) the experimental electric fields $E_r^{exp}(t)$ and $E_r^{ref}(t)$ of the sample and steel reference, respectively; and (c) $\varepsilon_1$ and $\sigma_1$ assumed for the analysis. Red arrows identify peaks corresponding to partial reflections.

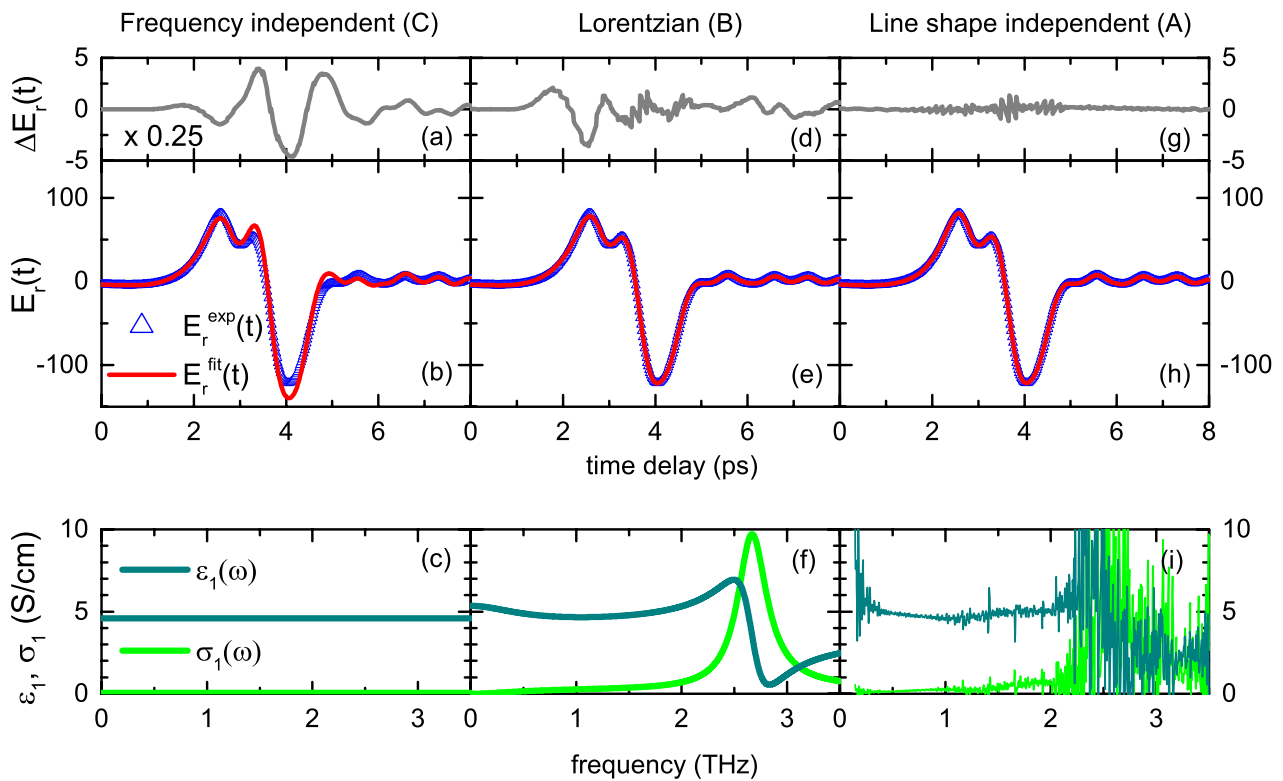

**Figure 2.** Fitting analysis schemes illustrated for $E_r^{exp}(t)$ of Fig. 1. The three schemes differ by the parametrization of the optical functions of the paint layer: (a-c) frequency independent $\varepsilon_1$ and $\sigma_1 = 0$; (d-f) summation of Lorentz oscillators; (g-i) variational dielectric function resulting in a quasi-line shape independence. (b,e,h,) Fits $E_r^{fit}(t)$ in the time-domain. (a,d,g) $\Delta E_r(t) = E_r^{fit}(t) - E_r^{exp}(t)$. (c,f,i) Optical functions of the paint layer resulting from the analysis procedures. Note that the noise in (i) is a direct consequence of the noise on the experimental data, pointing to lower signal-to-noise ratios in the frequency extremes of the spectra.

Most simple is to use the Fresnel equations for a thin slab on a substrate, using the optical properties as assumed for peak subtraction. The model takes into account all partially reflected and transmitted rays and visual peak identification is no longer required. The approximate optical properties, however, causes the fit $E_r^{fit}(t)$ to be of poor quality, as shown by the deviation $\Delta E_r(t) = E_r^{fit}(t) - E_r^{exp}(t)$ (Figure 2a).

Diverse light-matter interactions induce frequency dependence of the optical functions. Absorption and dispersion can be implemented in the model, e.g., by parameterizing the optical functions, where the parameters are to be adjusted during the fitting procedure. A practical example is a summation of Lorentzians, each representing a specific excitation. Figure 2d and 2e show that adding dispersion substantially improves $\Delta E_r(t)$ as compared to the previous scheme (Figure 2a). The optical functions obtained from the fitting procedure show characteristic signatures of the paint. The upturn of $\varepsilon_1$ corresponds to the GHz dielectric relaxation of water and the resonance around 2.6 THz to the typical boson peak. The center frequency, spectral weight and width of such features directly relate to the previously mentioned quality parameters of the constituents.

## Variational dielectric function

Although a Lorentzian line shape is often a good approximation, spectral features have their own, complex, line shape that is different from simple mathematical functions. This is very important for an accurate determination of their spectral weight.

Parameterization using a variational dielectric function allows for a quasi-line shape independent analysis scheme as shown in Figure 2h. The fit is almost perfect, as manifested by $\Delta E_r(t)$. Figure 2i shows the natural line shape obtained from the experimental data. In this procedure even Kramers-Kronig consistent noise is inherently taken into account. Although noise may be repelling as compared to smooth lines of Figure 2f, it shows a more correct result.

## Data analysis schemes compared

The quality of the two peak-subtraction and three fitting schemes is mutually compared in **Figure 3**. Three parameters are extracted there where possible: the deviation $|\Delta E_r(t)|^2$ integrated in the displayed time window (Figure 3a); the thickness $d$ as compared to conventional techniques (Figure 3b); and the spectral weight of the boson peak (Figure 3c). Peak subtraction assumes unknown optical properties, resulting in large deviations of $d$. The fitting approaches show that the closer the light-matter interaction is described by the scheme, the more and better material characteristic information can be extracted. Inclusion of dispersion and realistic line shapes are indispensable to obtain accurate spectral and geometrical properties.

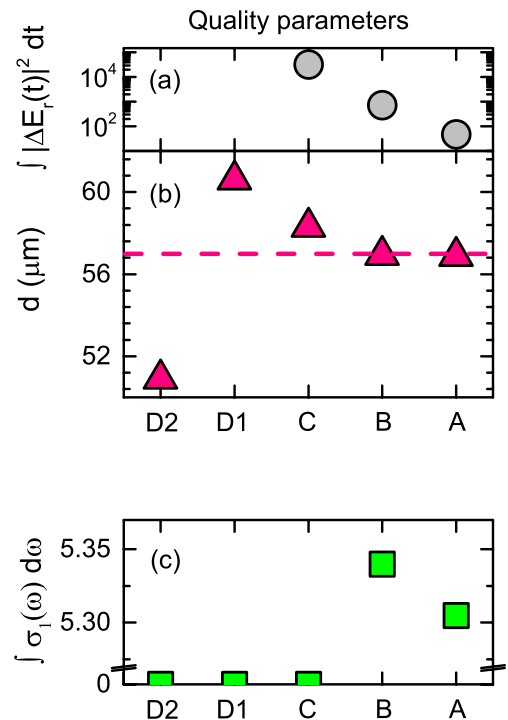

**Figure 3.** Quality parameters obtained from the peak subtraction (D2, D1) (Fig. 1) and fitting procedures (C, B, A) (Fig. 2). (a) Deviation $|\Delta E_r(t)|^2$ integrated in the time window of Fig. 2. (b) Paint layer thickness $d$ as compared to conventional methods (dashed line). (c) Integrated spectral weight of the boson peak as shown in Fig. 2f,i.

---

Dook van Mechelen is a physicist leading the THz technology integration at ABB Corporate Research, Switzerland.

## References

A.B. Kuzmenko. "Kramers–Kronig constrained variational analysis of optical spectra," Rev. Sci. Instr. **76**, 6849 (2005).

T. Yasui, T. Yasuda, K. Sawanaka, T. Araki. "Terahertz paintmeter for noncontact monitoring of thickness and drying progress in paint film," Appl. Opt. **44**, 6849 (2005).

S. Krimi, J. Klier, M. Herrmann, J. Jonuscheit, R. Beigang. ""Inline multilayer thickness sensing by using terahertz time-domain spectroscopy in reflection geometry," 38[th] Int. Conference on Infrared, Millimeter, and Terahertz Waves (2013).

J.L.M. van Mechelen, A.B. Kuzmenko, H. Merbold. "Stratified dispersive model for material characterization using terahertz time-domain spectroscopy," Opt. Lett. **39**, 3853 (2014).